\documentclass[conference]{IEEEtran}
\IEEEoverridecommandlockouts
\usepackage{orcidlink}
\usepackage{cite}
\usepackage{amsmath,amssymb,amsfonts}
\usepackage{algorithmic}
\usepackage{graphicx}
\usepackage{float}
\usepackage{textcomp}
\usepackage{xcolor}
\hypersetup{
	colorlinks,
	linkcolor={red!50!black},
	citecolor={blue!50!black},
	urlcolor={blue!80!black}
}
\def\BibTeX{{\rm B\kern-.05em{\sc i\kern-.025em b}\kern-.08em
    T\kern-.1667em\lower.7ex\hbox{E}\kern-.125emX}}
\begin{document}

\title{A Border Gateway Protocol Extension for Distributing Endpoint Identifier Reachability Information in Delay-tolerant Networks
\thanks{The intermediate results presented in this paper have been created in the context of the DARKSOL project. This project is co-funded by the European Union and co-financed from tax revenues on the basis of the budget adopted by Saxon State Parliament.}
}

\author{\IEEEauthorblockN{Marius Feldmann}
\IEEEauthorblockA{
\textit{D3TN GmbH}\\
Dresden, Germany \\
marius.feldmann@d3tn.com}
\and
\IEEEauthorblockN{Théo Tchilinguirian \orcidlink{0009-0000-5650-2446}}
\IEEEauthorblockA{
\textit{D3TN GmbH, Dresden, Germany }\\
\textit{Département Informatique, Polytech, Université Montpellier}\\
theo.tchlx@proton.me}
\and
\IEEEauthorblockN{Felix Walter \orcidlink{0000-0002-4724-8092}}
\IEEEauthorblockA{
\textit{D3TN GmbH}\\
Dresden, Germany \\
felix.walter@d3tn.com}
}

\makeatletter
\def\ps@IEEEtitlepagestyle{%
  \def\@oddfoot{\mycopyrightnotice}%
  \def\@evenfoot{}%
}
\def\mycopyrightnotice{%
  \begin{minipage}{\textwidth}
  \centering \scriptsize
  Copyright~\copyright~2025 IEEE. Personal use of this material is permitted. Permission from IEEE must be obtained for all other uses, in any current or future media, including reprinting/republishing this material for advertising or promotional purposes, creating new collective works, for resale or redistribution to servers or lists, or reuse of any copyrighted component of this work in other works.
  DOI: \href{https://doi.org/10.1109/WiSEE57913.2025.11229835}{10.1109/WiSEE57913.2025.11229835}
  \end{minipage}
}
\makeatother

\maketitle

\begin{abstract}
The Delay-Tolerant Networking (DTN) community has created solid results during the last three decades. One aspect that still requires focus, however, is the simplification of configuring systems participating in DTN communication. In this workshop paper, an approach is proposed for configuring DTN endpoint reachability. The core idea is to extend the Border Gateway Protocol (BGP), enabling it to advertise the reachability of DTN endpoints specified by Endpoint Identifiers (EIDs) via specific next-hop Convergence Layer Adapters (CLAs). The approach is mainly intended for usage in edge nodes accessible from an IP network, which communicate via a DTN gateway to a larger DTN infrastructure. It may be applied to simplify access to the Solar System Internet from hosts connected to the Internet. The feasibility of the approach has been confirmed using an implementation intended for validation purposes.
\end{abstract}

\begin{IEEEkeywords}
Delay-Tolerant Networking, Disruption-Tolerant Networking, Border Gateway Protocol, Endpoint Identifier Reachability
\end{IEEEkeywords}

\section{Introduction}
Delay-tolerant networks rely on tailored protocols that are capable of storing packets on individual hosts or routers until a suitable next hop is reachable. The applied store-and-forward approach is key for moving packets via paths of links without continuous availability and with long delays. The Bundle Protocol (BP) described in RFC 5050 (version 6) and RFC 9171 (version 7) is based on this principle. 

Several DTN protocol implementations such as the Interplanetary Overlay Network (ION), High-Rate Delay Tolerant Networking (HDTN), or µD3TN have emerged in the last two decades, using the Bundle Protocol. These implementations typically come with their own configuration interfaces and tools for handling aspects such as assigning endpoint identifiers (EID) to a Bundle Protocol Agent, configuring contacts to neighboring nodes, and making routing information available, e. g., in the form of a contact plan. 

Although the given configuration mechanisms are valid for various application contexts, they require additional effort to provide a Bundle Protocol Agent with all information necessary to perform its task. Furthermore, they typically assume that most of the information is available a priori. Although solutions have been developed for dynamically discovering DTN neighbors and for sharing reachability information among them, these approaches focus on dynamic ad-hoc networks in which multicast or even broadcast communication is common for advertising reachability information. 

Simplifying the configuration of a Bundle Protocol Agent is essential to promote usage of the Bundle Protocol in challenged networks. Accessing a delay-tolerant network should be as simple as accessing the Internet. For this to be the case, mechanisms are required that automatically hand over all necessary configuration information to a Bundle Protocol Agent. Developing holistic solutions in this context is a huge open task for the DTN community.

With the proposal described in this paper, we would like to contribute a small but essential item to the solution space. Our concept allows for automating the distribution of reachability information in DTNs as soon as a link to a neighbor is active. This makes it possible to launch a Bundle Protocol Agent without additional configuration of reachable peers, which then automatically receives information about which EID is currently reachable via which CLA. In order to integrate with existing routing software implementations, the Border Gateway Protocol is used as the basis for the proposed approach.

To begin with, we provide an overview of two relevant use cases that could benefit from the proposed solution in section \ref{sec:usecases}. Secondly, we discuss the state of the art of providing reachability information in DTNs in section \ref{sec:sota}. Afterwards, the proposed solution is described and a roadmap for extending its application domain is sketched in section \ref{sec:concept}. A prototype of the BGP-based reachability information distribution approach named \textit{Hermes} that has been developed for validation purposes is addressed in section \ref{sec:prototyping}. The paper concludes with a summary and an outlook on future work in section \ref{sec:summary}.

\section{Use Cases}\label{sec:usecases}

In order to motivate the proposed approach, we consider two use cases for applications.

\subsection{Access to the Solar System Internet}

The vision of a Solar System Internet (SSI) is on the roadmap of several space agencies. Specifically, the European Space Agency (ESA) has clearly stated the relevance of the SSI in a Vision 2040 document\footnote{\url{https://esamultimedia.esa.int/docs/technology/Technology_2040.pdf}}. Delay-Tolerant Networking protocols are assumed to play an essential role in implementing the SSI.

As it is simple to participate in the regular Internet, simple accessibility should be a key focus for the SSI to promote its usage. In ideal circumstances, it becomes possible to start a DTN node in a terrestrial IP-based edge network region of the SSI, which then automatically receives all configuration and reachability information required to participate in the SSI. In the considered case, part of this information has to define reachable EIDs plus associated next-hop CLA endpoints. The approach introduced in section \ref{sec:concept} is intended to be applicable in this specific context.

\subsection{Delay-tolerant Kubernetes}

Together with two collaboration partners, Cyberus Technology and Dresden University of Technology, D3TN is participating in the R\&D project DARKSOL\footnote{\url{https://darksol.cloud/}}, which aims to investigate how application execution platforms such as Kubernetes could be deployed and operated via a DTN infrastructure. Each instance of the platform itself is deployed in a network segment with continuously available links and low delays. However, the infrastructure supporting the platform, the platform software (e. g. Kubernetes), and applications deployed on the platform have to be managed and accessible through delay-tolerant networks.

One of the requirements addressed by the project is that new DTN endpoints may be registered for applications executed on such a platform at any point in time. De-registration must also be supported and the set of running applications can change dynamically.
The network beyond the Kubernetes cluster that executes such an application has to be aware of these changes in the set of reachable endpoints. Therefore, a mechanism for distributing EID reachability information is required. 

\section{Requirements and State of the Art}\label{sec:sota}
In order to clearly delineate the focus of the proposed approach, a set of eight core requirements (termed R1-R8) has been identified that are addressed by the BGP extension described in section \ref{sec:concept}. These requirements can be derived from the aforementioned use cases. They are summarized in the following section \ref{subsec:requirements} before an overview of the state of the art is provided in section \ref{subsec:sota}.

\subsection{Requirements}\label{subsec:requirements}

The core focus of the proposed protocol mechanism is to bridge the gap from the Internet to delay-tolerant networks of heterogeneous technological kind. Thus, it is essential that the approach is capable of operating in IP networks and must be able to distribute reachability information through these networks (\textit{R1}). 

Secondly, information exchange must be carried out based on an existing protocol that is applicable in IP networks to reduce the implementation effort and to allow reuse of operational knowledge (\textit{R2}). 

The approach must be capable of communicating to a neighboring node that a BP endpoint or a set of endpoints is currently reachable via a specific next hop, and capable of invalidating this reachability information (\textit{R3}). To emphasize the scope of the proposed approach, it is explicitly pointed out that it is not intended to distribute information about planned communication opportunities (i. e., contacts) represented in a contact plan. Instead, it can communicate from a node B to a neighboring node A that, \textit{at the current time}, bundles destined for a specific EID can be handed over from node A to node B, as node B offers a path to the endpoint identified by this EID. Likewise, it can invalidate this reachability information, e. g., node B may tell neighboring node A that, \textit{at the current point in time}, it cannot reach a specific EID or a set of EIDs anymore.

Different Convergence Layer Adapters (CLA), such as the TCP Convergence Layer Version 4 (TCPCLv4), the UDP Convergence Layer (UDPCL) Version 2, or in the future potentially a QUIC-based CLA, may be used. Due to this, the approach shall be generic in relation to the CLA parameters communicated with the exchanged reachability information (\textit{R4}).

Likewise, different EID schemes must be supported (\textit{R5}). There shall be no additional constraints for the EID schemes used beyond to the extend to which they are supported by the Bundle Protocol.

Not only the endpoints registered at direct neighbors shall be communicated using the approach. It shall support the transitive distribution of the EID reachability information (\textit{R6}).

In addition to transmitting EID and CLA information to neighbors, the approach must be able to transfer auxiliary attributes of the endpoints (\textit{R7}). An attribute may be, for example, a cryptographic public key bound to the EID that can be used to exchange session keys to establish a security context. 

Finally, the EID reachability distribution protocol shall not be limited to a single implementation of the BP. It must offer simple mechanisms for integration with existing BP implementations (\textit{R8}) to facilitate adoption.

The described requirements represent the basis of the proposed approach outlined in section \ref{sec:concept} and the prototypical implementation presented in section \ref{sec:prototyping}.

\subsection{State of the Art}\label{subsec:sota}

As our intent is to bridge the gap between networks based on IP and BP, foundational and related work exists in both domains. Firstly, in IP-based networks, routing protocols are used for exchanging topological information. Secondly, in DTNs, there are several dynamic discovery mechanisms as well as techniques for routing information distribution.
To the best of our knowledge, no existing approach unifies the distribution of DTN reachability information with an IP routing protocol. Thus, this discussion focuses on the two areas individually.

A core objective of this work is to distribute DTN-specific reachability information in and through IP-based networks. In this domain, there are various protocols available for exchanging topological (routing) information.
These mechanisms are usually classified either based on their intended application area (e. g., Interior or Exterior Gateway Protocols) or depending on the type of information they distribute (e. g., distance-vector, link state, or path vector routing protocols).

IP-based techniques can fulfill requirements R1 and R2, as they natively operate in an IP environment. They however need to be extended to support any DTN-specific functionality -- no DTN integrations are available in state-of-the-art IP routing protocols at the time of writing.
Thus, a central criterion for our work was the extensibility of the chosen protocol and the available implementations.

RIP (RFC 1058 / RFC 2453) is a very simple protocol. However, due to the fixed set of header fields and the payload size limit of 512 bytes, it is not suitable for the intended application domain.
Other IGPs, like IS-IS (ISO/IEC 10589:2002) and OSPF (RFC 5340), support custom fields based on a type-length-value (TLV) representation and can also propagate them transitively.
These two protocols exchange link states with peers, making it possible to perform route selection with multiple concurrent alternatives, but also carry a higher overhead due to the increased amount of information that is synchronized throughout the network.
While still being comparably simple and having less overhead due to not distributing the full topology view like link-state protocols, BGP (RFC 4271) supports optional path attributes and Network Layer Reachability Information (NLRI) with an extensible set of types, both of which can be of variable length, offering potential for extensions.


In the context of DTNs, various techniques have been developed to spread topological information.
DTN IP Neighbor Discovery (IPND) \cite{ipnd} transmits UDP beacon messages advertising the DTN node's EID and available CLA addresses via an IP-based underlay and can be used to derive presence and reachability information concerning direct neighbors in such a network.
Ring Road Neighbor Discovery (RRND) \cite{rrnd} presents a generalization of IPND that, in extension to removing the requirement to use UDP and introducing additional security measures, also transmits a list of transitively reachable EIDs, although it still only operates on a single-hop basis.
BPv7 Secure Advertisement and Neighborhood Discovery (SAND) \cite{sand}, a recent IETF draft, leverages the Bundle Protocol to transmit its messages and thus can use BPSec for security. Like IPND and RRND, SAND advertises a next possible DTN hop along with its CLA interfaces and additional properties. This protocol can be considered the most recent and complete approach for DTN neighbor discovery.

For informing other DTN nodes about transitively reachable peers, several solutions exist, which depend on the used routing scheme. While opportunistic and probabilistic DTN routing mechanisms usually define an own information exchange procedure (e. g., PRoPHET, defined in RFC 6693, which distributes delivery predictability values and aggregates them on each intermediary hop), deterministic routing (with Contact Graph Routing \cite{sabr} and its variants being the most prominent and widely used examples) often considers this issue orthogonal to the actual routing algorithm. Thus, a static configuration can be used that can be updated through DTN management facilities, or a mechanism can be deployed to distribute contact plans throughout the network, such as the Contact Plan Update Protocol (CPUP) \cite{cpup}.

It becomes evident that the mechanisms available from the DTN domain either solely focus on next-hop reachability (i.e., are \emph{discovery} approaches) or require the distribution of transitive information concerning the full time-varying network graph and its characteristics (e. g., contact time intervals or expected recurrence probabilities).
A simple transitive route exchange and update mechanism for IP-based network segments of an overall DTN with less volatile link characteristics is not available at the time of writing.

\section{EID Reachability Information Distribution}\label{sec:concept}
\begin{figure*}[h]
    \centering
    \includegraphics[width=0.85\textwidth]{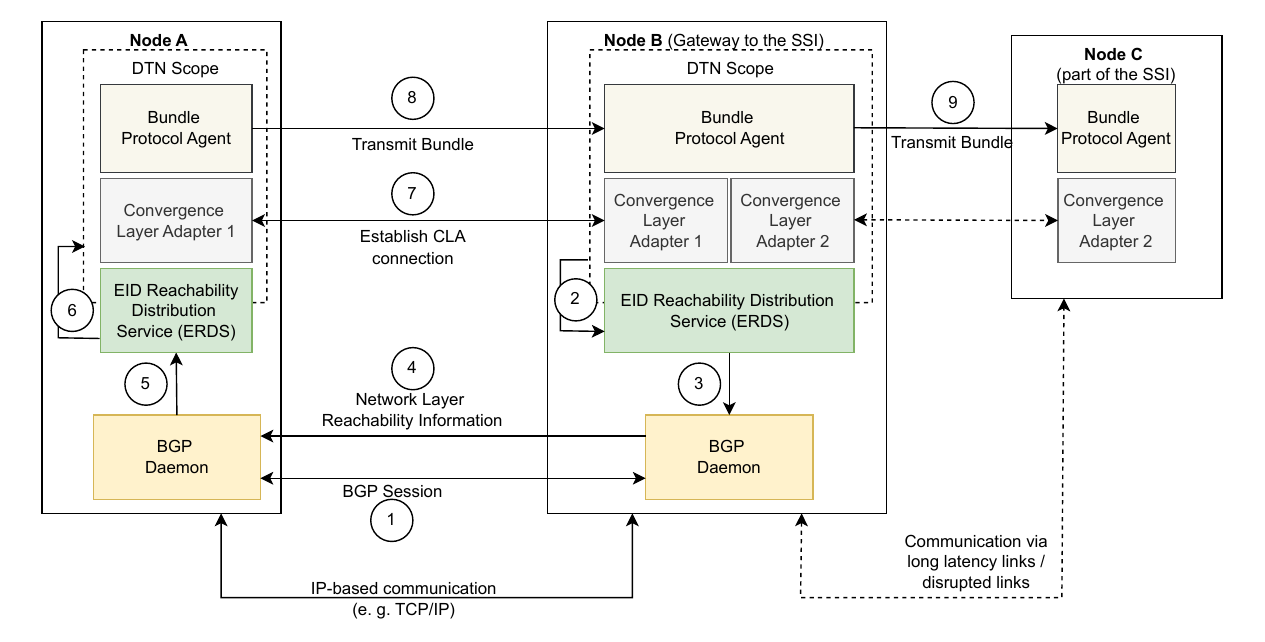}
    \caption{Global Picture of the EID Reachability Distribution Approach: Two nodes connected via a low-latency IP communication channel exchange EID reachability information using BGP. A third node, connected via a DTN link, is announced in the IP subnetwork but does not participate itself in the BGP-based reachability information distribution scheme. The individual steps for EID reachability distribution are labeled in the figure.}
    \label{fig:global-picture}
\end{figure*}

After providing an overview of the state of the art, we introduce the proposed approach in the next step.
Firstly, the use of BGP is discussed. Secondly, the anticipated information exchange procedure is sketched. Finally, the focus is on the extension of the Border Gateway Protocol that enables the transmission of EID and CLA information.

\subsection{Using BGP for Distributing EID Reachability Information}\label{subsec:bgp}

As BGP is used intensively in IP networks to distribute relevant topological information for overlay networks such as Ethernet Virtual Private Networks (EVPN, see RFC 7432), it has been selected as a basis for the designed EID reachability information distribution approach. Because it operates via the Internet Protocol, it also meets requirement R1. In addition, it offers a basic information distribution mechanism for exchanging reachability information between peers. Thus, its usage meets requirement R2, as no dedicated message-exchange mechanism has to be designed from scratch. Leveraging BGP also addresses requirement R3 -- owing to the BGP Multiprotocol Extension, the attributes \texttt{MP\_REACH\_NLRI} and \texttt{MP\_UNREACH\_NLRI} have been introduced in RFC 4760. These attributes make it possible to point to a next hop that is reachable via a particular network layer protocol and to share information about protocols on layers above this network layer (including overlay networks). Thus, EID reachability information can be transmitted to neighbors and can be updated and invalidated with these attributes.

In order to make the approach work, two contributions to the state of the art have to be made:
\begin{enumerate}
    \item A dedicated NLRI must be specified to distribute the EID reachability information. Support for this NLRI has to be added to implementations of BGP.
    \item A service managing the information exchange between a Bundle Protocol Agent and a BGP daemon has to be developed and integrated in a lean manner on both sides.
\end{enumerate}

In subsection \ref{subsec:nlri} a draft for the DTN-specific NLRI is described. A proof of concept for the implementation-related contributions is sketched in section \ref{sec:prototyping}.

In addition to the minimally required contributions to create an EID reachability distribution approach, further work may extend the applicability of the proposed mechanism. Currently, the IETF IDR WG works on an approach to operate BGP over QUIC. Furthermore, the IETF TIPTOP WG works on ``(r)ecommendations for the configuration and deployment of QUIC (and associated security protocols) for space networking.''\footnote{\url{https://datatracker.ietf.org/doc/charter-ietf-tiptop/01/}} Combining these two efforts and transferring them to the mechanism proposed by us would make it possible to spread not only reachability information via the Internet but also via DTN links. However, extending the approach into this area is beyond the scope of this contribution.

\subsection{Information Exchange Approach}

The general EID reachability information exchange approach is described on the basis of a simple scenario shown in Fig. \ref{fig:global-picture}:
Node A is connected through the Internet to Node B, which forms the gateway to a delay-tolerant network with long-latency or disrupted links. For example, Node B could be a gateway operated in a ground station for transferring bundles via the SSI to nodes located on other celestial bodies. 

When Node A starts up, it establishes a BGP session to Node B (step 1).
Every node participating in the information exchange executes an \emph{EID Reachability Distribution Service (ERDS)}, which receives all updates of reachable EIDs and CLA parameters from the local Bundle Protocol Agent (step 2). On Node B, a trigger for this may be that a new EID has appeared in a Contact Plan update related to Node C.

The EID and CLA information is handed over to the local BGP daemon (step 3) and, afterwards, transferred via a BGP message leveraging the \texttt{MP\_REACH\_NLRI} attribute to Node A (step 4). Its BGP daemon forwards the reachability information through the ERDS to the BP implementation (steps 5 and 6), which uses it to configure its Bundle Protocol Forwarding Information Base as well as its Convergence Layer Adapters. For forwarding bundles to the previously received EID, the node establishes a CLA connection (e. g. a TCP connection) using the previously received parameters (step 7). Afterwards, a bundle destined for the EID received via BGP is forwarded to Node B (step 8). This bundle is then handed over (based e. g. on a common DTN routing approach) from Node B to Node C (step 9) from where it may be transmitted further within the Solar System Internet. 

The scenario shows how transitively reachable EIDs are distributed with the proposed mechanism. Thus, fulfillment of requirement R6 is confirmed. As the ERDS is used as a glue component between the BP implementation and the BGP daemon and can be used for different implementations of both, the approach addresses requirement R8.

The discussed approach allows Node A to participate in Bundle Protocol communication without the need for manual configuration or additional contact exchange mechanisms. The only required configuration is that of a BGP peer that provides the necessary EID reachability information, which could be either hard coded or provided through the Domain Name System (DNS).

\subsection{Multiprotocol Reachable and Unreachable NLRI}\label{subsec:nlri}
As mentioned in section \ref{subsec:bgp}, specific NLRI have to be introduced in order to equip BGP with the capability to distribute EID reachability information and later invalidate this information. The generic structure of these attributes as displayed in RFC 4760 on pages 3 and 5 has been adapted to represent the information required for the exchange of EID reachability information. 
The structure of the used \texttt{MP\_REACH\_NLRI} is sketched below:

\begin{small}
\begin{verbatim}
       +------------------------------+
       | AFI (2 octets) = 23042       |
       |------------------------------|
       | SAFI (1 octet) e.g. 0 = MTCP |
       +------------------------------+
       | Length of NHNA  (1 octet)    |
       +------------------------------+
       |   Next Hop NA   (variable)   |
       +------------------------------+
       | Number of NLRI  (1 octet)    |
       +------------------------------+
       |  NLRI/EID Data  (variable)   |
       |  +------------------------+  |
       |  | URI Code    (1 octet)  |  |
       |  |  (e.g. IPN, DTN...)    |  |
       |  +------------------------+  |
       |  | EID Length  (1 octet)  |  |
       |  +------------------------+  |
       |  | EID Value   (variable) |  |
       |  |  (e.g., ipn:5.1)       |  |
       |  +------------------------+  |
       |  | No. of Attrs. (1 octet)|  |
       |  +------------------------+  |
       |  | Attr. Data   (variable)|  |
       |  |   Type-Length-Values   |  |
       |  +------------------------+  |
       +------------------------------+
\end{verbatim}
\end{small}

The following values assigned to the different fields of the attribute are just exemplary. The assignment of final values is a task for a potential future standardization activity.

For the purpose of validating the approach, the value \texttt{23042} has been assigned to the \emph{Address Family Identifier (AFI)}. 

The value of the \emph{Subsequent Address Family Identifier (SAFI)} depends on the CLA used. For example, the value \texttt{0} can be leveraged for the Minimal TCP Convergence-Layer Protocol (MTCP), \texttt{1} for version 3 of TCPCL, and so on. 

The \emph{Next Hop Network Address (NHNA)} contains the CLA endpoint of the next-hop DTN node. If this consists of an IP address plus a UDP or TCP port, these values are concatenated. Thus, the binary representation of a 128-bit IPv6 address concatenated with a 16-bit TCP port number may be used to point to an MTCP CLA endpoint of the next hop. In this case, the field \emph{Length of Next Hop Network Address (Length of NHNA)} contains the value \texttt{144}.

The field \emph{Number of NLRI} provides the number of EIDs advertised. This field is followed by information about the EID(s). For each EID, the URI Code (e. g. the value \texttt{1} for the \texttt{dtn} EID scheme), the EID length, and the EID value are provided. The EID is encoded as described in the Bundle Protocol specification. Additional data, such as the next hop's node identifier or public keys associated with the EIDs, may be provided. For this purpose, the field \emph{Attr. Data} can be used. The value of the field \emph{No. of Attrs.} represents the number of additional attributes.

The discussed NLRI information addresses requirements R4, R5, and R7.

Invalidation of EID reachability information is done using a similar \texttt{MP\_UNREACH\_NLRI} attribute that lists in its \emph{Withdrawn Routes} field the EIDs that are not reachable anymore. For this purpose, the EID information can be encoded in the same manner as in the case of \texttt{MP\_REACH\_NLRI}.

\section{Prototyping}\label{sec:prototyping}

In this section, we present a software suite for BGP-based distribution of EID reachability information that is used to validate the proposed EID reachability exchange mechanism. It consists of µD3TN as BP implementation, BIRD (extended with DTN support) as BGP implementation, and \emph{Hermes}, a new software implementing an ERDS. BIRD's implementation of BGP was chosen as it supports Multiprotocol Extensions. µD3TN was selected thanks to its AAP2 interface that simplifies the exchange of forwarding information. 

\subsection{Overview of Implementation}

\emph{Hermes} is a Rust program that connects to exactly one BP and one BGP implementation, with the purpose of receiving EID reachability information from either of the two and acting as a compatibility layer to translate and forward it to the other.

\begin{figure}[htbp]
  \centering
  \includegraphics[width=0.48\textwidth]{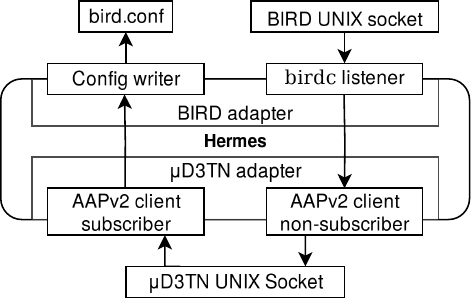}
  \caption{Hermes high-level structure: The modular architecture has adapters in both directions, for integrating the BGP and BP implementations.}
  \label{fig:hermes-arch}
\end{figure}

A key design property of Hermes is its straightforward extensibility. Enabling support for more BP or BGP implementations is done by writing \emph{adapters} following a common software interface to translate and exchange EID reachability information. Fig. \ref{fig:hermes-arch} illustrates the internals of an instance connected to µD3TN and BIRD.

Currently, Hermes implements adapters only for µD3TN and BIRD. The adapter interface defines functions to send and listen for EID reachability information. Adapters can implement such functionality as needed. EID reachability information is represented internally as strongly typed data structures to ensure the interface is compatible between adapters, and in turn between any BP and BGP implementations supported through adapters.

\subsection{Scenario-based Validation}

\begin{figure}[htbp]
  \centering
  \includegraphics[width=0.38\textwidth]{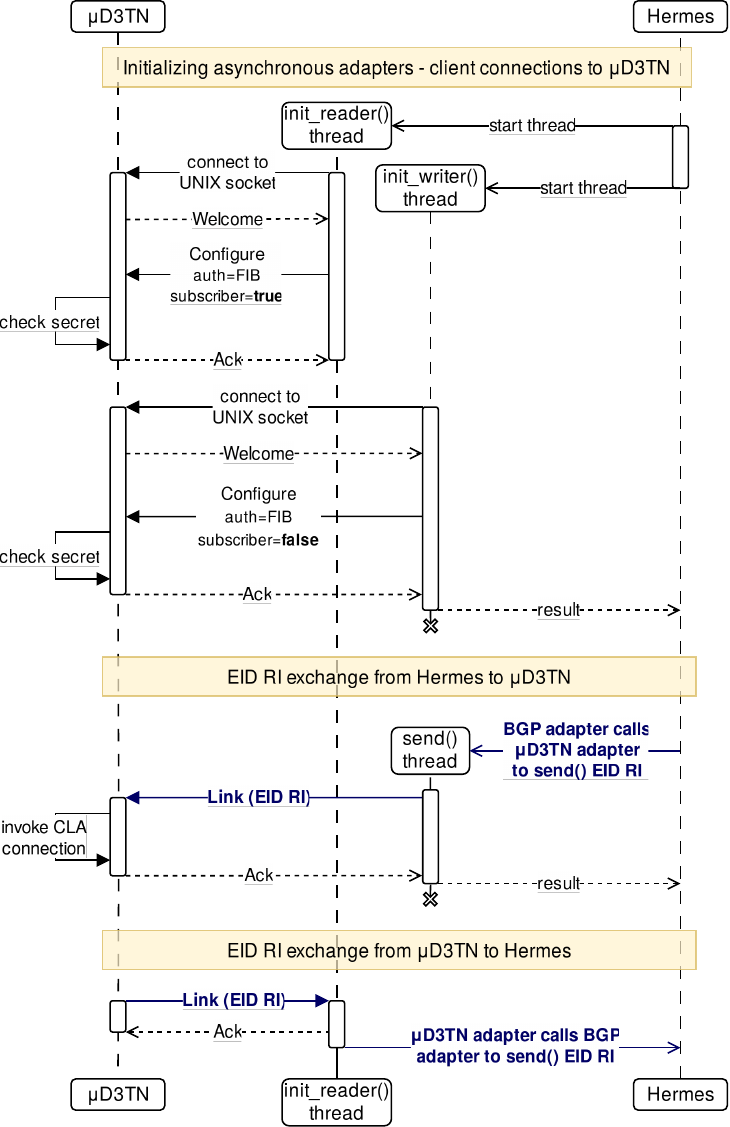}
  \caption{EID reachability information exchange between µD3TN and Hermes. The AAP2 interface is used to perform updates on demand in both directions.}
  \label{fig:hermes-sequence}
\end{figure}

BP and BGP adapters are specified and configured in a TOML configuration file. Fig. \ref{fig:hermes-sequence} shows the initialization of the adapter listener and sender in separate threads, as well as the exchange of EID reachability information (abbreviated as EID RI) between µD3TN and Hermes on a single host. µD3TN AAP2 Protobuf messages are capitalized. 
The BIRD adapter is not depicted as it is much less complex: It does not require connection establishment, simply calling the \verb|birdc| command asynchronously to retrieve the EID reachability information and writing or deleting static routes to propagate it.

\begin{figure}[htbp]
  \centering
  \includegraphics[width=0.4\textwidth]{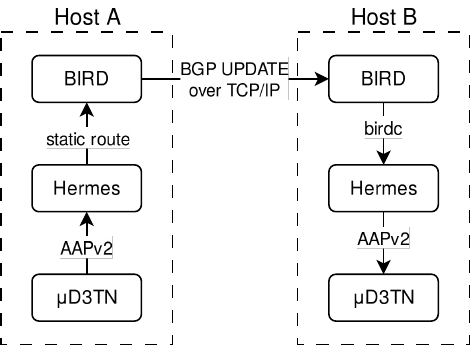}
  \caption{BGP exchange between two DTN-enabled hosts}
  \label{fig:hermes-deployment}
\end{figure}

Fig. \ref{fig:hermes-deployment} shows an exchange of EID reachability information from host A to host B, through Hermes adapters between µD3TN and BIRD.
Current deployments of this implementation have demonstrated the feasibility of the proposed approach, by transferring EID reachability information from a µD3TN-specific format to the draft \texttt{MP\_REACH\_NLRI} format for propagation through BGP with an extended version of BIRD.


\section{Summary and Outlook}\label{sec:summary}
In this workshop paper, we have motivated an EID reachability information distribution protocol and provided an overview of the state of the art in this context. Based on a discussion of the requirements, it has been pointed out that it is explicitly not intended to distribute time-varying topology information like contact plans. Instead, we focus on making it possible to advertise EID reachability via available next hops in a DTN.

As a basis for the proposed protocol, BGP has been chosen, which is particularly suitable because of its extensibility. BGP makes it possible to transfer NLRI reachability information and relevant updates to BGP neighbors. Using this mechanism, a dedicated protocol extension has been proposed in section \ref{sec:concept}. As mentioned, the distribution mechanism fulfills the identified requirements R1 to R8.

The approach was implemented\footnote{\url{https://github.com/darksol-cloud}} for validation purposes, using the BGP implementation of BIRD as well as µD3TN with its AAP2 interface. Based on this implementation, the concept has been fundamentally validated.

In future work, the draft implementation will be extended and made available as an open source tool to the community. Additionally, it is planned to compare the BGP-based approach to one that uses the Intermediate System to Intermediate System (IS-IS) routing protocol. Furthermore, a BGP variant that uses QUIC instead of TCP may be explored.

If the approach is seen as relevant by the DTN community, standardization activities in IETF's DTN WG will be initiated.


\begin{thebibliography}{00}





\bibitem{ipnd}
	D. Ellard et al.,
	``DTN IP Neighbor Discovery (IPND),''
	\emph{Internet-Draft},
	IETF, 2023.
	\url{https://www.ietf.org/archive/id/draft-johnson-dtn-ipnd-00.html}
	
\bibitem{rrnd} 
	M. Feldmann and F. Walter,
	``Towards Ground Station Contact Discovery in Ring Road Networks,''
	in \emph{IEEE International Conference on Wireless for Space and Extreme Environments (WiSEE)}, Orlando, FL, 2015.

\bibitem{sand}
	B. Sipos and J. Deaton,
	``Bundle Protocol (BP) Secure Advertisement and Neighborhood Discovery (SAND),''
	\emph{Internet-Draft},
	IETF, 2025.
	\url{https://datatracker.ietf.org/doc/draft-ietf-dtn-bp-sand/01/}

	
\bibitem{sabr}
	``Schedule-Aware Bundle Routing,''
	\emph{CCSDS 734.3-B-1}, Washington, DC, USA, 2019.
	
\bibitem{cpup}
	N. Bezirgiannidis et al.,
	``Towards flexibility and accuracy in space DTN communications,''
	in \emph{Proceedings of the 8th ACM MobiCom workshop on Challenged networks (CHANTS '13)}
	ACM, New York, NY, USA, 2013.

\end{thebibliography}
\end{document}